\documentclass[12pt]{article}

\usepackage[utf8]{inputenc}
\usepackage[T1]{fontenc}

\usepackage{amsfonts}
\usepackage{nicefrac}

\usepackage[margin=1in]{geometry}
\usepackage{setspace}
\usepackage{natbib}
\setcitestyle{authoryear,open={(},close={)}}

\usepackage[
  colorlinks=true,
  linkcolor=black,
  citecolor=black,
  urlcolor=blue,
  breaklinks=true
]{hyperref}
\usepackage{url}

\usepackage{booktabs}
\usepackage{graphicx}
\usepackage{subcaption}
\usepackage[export]{adjustbox}
\usepackage{float}
\usepackage{longtable}
\usepackage{rotating}
\graphicspath{ {./images/} }

\usepackage{chngcntr}

\usepackage[hang,flushmargin]{footmisc}

\usepackage{enumitem}

\usepackage{lipsum}
\usepackage{color}
\usepackage{fancyhdr}
\usepackage{silence}
\makeatletter
\renewcommand{\maketitle}{%
  \begin{center}
    {\large\bfseries \@title \par}
    \vskip 1.5em
    {\normalsize Lauren Kahn\par}
    {\small Center for Security and Emerging Technology, Georgetown University\par}
    \vskip 0.5em
    {\normalsize Michael C. Horowitz\par}
    {\small University of Pennsylvania\par}
    \vskip 0.5em
    {\normalsize Laura Resnick Samotin\par}
    {\small United States Military Academy\par}
    \vskip 0.5em
    {\normalsize \@date\par}
  \end{center}
  \vskip 1.5em
}
\makeatother
 
\title{What is Human in Judgment? Comparing Automation Bias and Algorithm Aversion Between the United States Military Academy and the General Public}
 
\date{\today}
 
\begin{document}
 
\maketitle

\clearpage

\begin{abstract}
Human judgment has always been central to conflict and escalation, but how will a world of artificial intelligence (AI) change the role of humans in war? As militaries increasingly adopt AI-enabled decision-support systems (DSS), including the United States in the war against Iran, concerns about automation bias---over-reliance on algorithmic recommendations---and algorithm aversion---premature distrust of automated outputs---raise fears that relying on AI too much could increase the risk of error, miscalculation, and accidents. Yet existing evidence on how militaries actually interact with AI remains limited. We test theories about the susceptibility of militaries to automation bias by comparing the results from a survey experiment conducted with 236 cadets at the United States Military Academy at West Point to a demographically similar cross-national public sample. Respondents completed a target identification task and then received advice from either an algorithm or a human analyst and had the opportunity to re-assess their initial identification, allowing direct measurement of automation bias and algorithm aversion. We find that West Point cadets are less prone to cognitive distortion than members of the general public, displaying better calibrated trust in algorithmic decision support systems. While the findings are limited, they suggest that military education and exposure to AI can meaningfully shape how AI influences international politics in matters of war and peace.
\end{abstract}


\newpage

\section{Introduction}

International relations theories of war and peace are underpinned by an understanding of human agency and the way human judgment informs decisions about conflict and escalation. Artificial intelligence (AI) could meaningfully alter traditional decision-making pathways for leaders and militaries in ways that impact politics. Understanding how AI will shape how human decision-making is therefore essential to accurately assessing the future of politics and conflict.

The need to grasp how AI will shape decision-making is acute since militaries worldwide are developing and fielding AI solutions from the boardroom to battlefield, integrating algorithms into areas including predictive maintenance, logistics, disaster relief, nuclear command and control, automatic target recognition in weapons systems, and robotics \citep{HicksCDAO, DigitalTargeting, stepanenko, lin2025, AIArmageddon, HumanMachineTeaming}. As militaries integrate AI technologies, they will, for the foreseeable future, still require commanders and operators to exercise human judgment. While militaries such as the United States have national policies to ensure human oversight of the use of force \citep{DoD3000.09, AppropriateLevels, IHLandAWS}, they are also leveraging AI Decision Support Systems (AI DSS) to speed up targeting processes by delivering information to commanders faster, as the US is in its war against Iran \citep{Manson}. Theories of cognitive bias unique to automated systems---automation bias and algorithm aversion---suggest, however, that military users may over-rely on or prematurely distrust automated outputs in ways that increase the risk of accidents or mistakes, which, in crisis or combat contexts, can have life-or-death consequences \citep{Ding2024/2025, LawfareRisks}. There is therefore a strong need to assess how and when militaries are at risk of introducing bias through AI decision-support tools.\footnote{If this is the case, then further aspects of decision-making---including at lower levels of command than international relations theories typically examine---should be taken into account when modeling the outcomes of future crises or wars.}

The combination of increasing autonomy, pervasive computerization, and the opaque, often black-box nature of modern AI systems makes automation bias a prominent concern across both civilian and military domains \citep{AIDSSLegalRisks, bode2024problem, garcia2024algorithms}. However, there is limited behavioral research on automation bias in military contexts, making it ripe for analysis.

This article compares findings from a survey experiment conducted in May 2025 of 236 cadets at the United States Military Academy at West Point (USMA), with those from a cross-national survey experiment of 702 members of the general public across 9 countries, including the United States \citep{horowitzkahnisq}. The experiment measures respondents' decision-making in response to assistance from an algorithm or a human analyst during a theoretical military target identification task. By comparing future military officers with members of the public similarly situated in age and education, we can gain analytical leverage on the extent to which automation bias and algorithm aversion are concerns for the military.

The results provide some evidence that the rising generation of Army officers, at least in the United States, is more appropriately calibrated to the upsides and risks of AI than demographically similar members of the general public. The training current West Point cadets receive in warfighting and high-consequence decision-making, including decision-making with AI, appear to make appropriate use more likely, though we cannot definitively prove that within the confines of this paper. The results suggest that there are opportunities for militaries to mitigate the negative predicted consequences of AI DSS on miscalculation, which would have substantial consequences for international politics, and require further investigation. In what follows, we lay out the background and theory, including the interest in AI DSS among militaries, discuss cognitive factors that influence human-algorithm interaction, present the research design and findings, and conclude by assessing the impact for international politics and future research.

\section{Automation Bias, International Politics, and AI Decision Support Systems}

International relations theories of war and peace rely on the notion of human decision-making. Whether considering escalation risk and the potential for conflict spirals or the use of military power, decisions by leaders, diplomats, military officers, and others drive continuity and change in world politics \citep{jervis1976perception, biddle2004military}. Existing research shows how many factors influence use of force decision-making, including institutional constraints \citep{TrumboreandBoyer, MontenandBennett}, group psychology \citep{JanisGroupthink, HerekJanisHuth, Kertzer_Holmes_LeVeck_Wayne_2022}, cognitive biases and rationality \citep{Williams, GrossStein2017}, educational and personal background \citep{BiddleMinSamotin}, and even emotions \citep{Zilincik, McDermott2026}.

However, the environment in which decision-makers---particularly military decision-makers in technical fields, such as target assessment---operate is changing as AI-driven technologies are introduced \citep{ONRonMilDecisions}. This raises the question of how human judgment over the use of force functions as AI and automated systems increasingly shape the battlefield and decision environments. AI DSS are a key potential use case for militaries applying AI on the battlefield. AI DSS consist of ``a class of algorithms that provide calculations to solve complex problems or improve task performance, and can suggest solutions that can either be accepted or rejected by a human expert'' \citep{ImpactsOfDSS}. AI DSS are already being used across sectors, from medicine to aviation. In military contexts, these systems are designed to leverage advances in sensing, data fusion, and machine learning to support human commanders in navigating complex battlefields \citep{HumanMachineWar, OODALoop}.

Most military AI DSS are still in the research and development phases, but some are now being fielded. Military AI DSS can currently assist with tasks including situational awareness, pattern detection and prediction, course-of-action generation, enabling functions such as identifying friendly and adversary forces, routing forces, allocating logistics, collecting intelligence, and prioritizing medical care \citep{ImpactsOfDSS, NotOracles}. In Israel's operations in Gaza, the Israeli Defense Force has controversially used AI DSS Gospel and Lavender to generate targeting recommendations and search extensive databases of suspected combatants.\footnote{The specific controversies surrounding Israel's use of AI DSS are beyond the scope of the paper. It appears, however, initial reporting that Lavender would generate potential targets and human operators would only have seconds to make decisions on a strike, increasing the chance of errors, was likely incorrect. Instead, the use of Lavender reflected the confidence levels accepted by human commanders, i.e. Israel's established rules of engagement \citep{Dwoskin}.} Likewise, in the Russia-Ukraine war, Ukraine uses AI DSS such as Delta, GisArta, and Kropyva to consolidate and query battlefield sensor data and provide algorithmically-derived recommendations \citep{UkraineMOD, UberForArtillery}.

The United States' Maven Smart System (MSS), named after the Pentagon's initial flagship AI initiative, Project Maven, uses computer vision, machine learning, and data integration to facilitate accurate target identification. Prior to 2026, American units such as the 18th Airborne Corps used MSS to reduce the time needed to accurately target by a factor of 100 with only twenty personnel, achieving performance comparable to Operation Iraqi Freedom's time-critical targeting cell, which required over two thousand support staff to function effectively \citep{ProbascoCSET}. In the US-Israel war against Iran, US Central Command reportedly used MSS extensively to speed the targeting cycle, quickly and automatically getting information to senior commanders for decision in minutes that used to take hours or days. AI DSS is a key part of how the U.S. has been able to attack thousands of Iranian targets in a short period of time \citep{Manson}.

Military interest in adopting AI DSS raises questions about their appropriate use given the risks of automation bias and algorithm aversion. Automation bias is the ``tendency [for human operators] to over-rely on automation'' \citep{Goddard2012}. Individuals exhibiting automation bias demonstrate cognitive offloading, or overdelegation to automated systems, reducing active monitoring of both system performance and the task at hand relative to what is warranted given the accuracy of the technology or importance of the task. In doing so, they place disproportionate trust in the system's recommendations, treating them as a ``heuristic replacement of vigilant information seeking and processing'' \citep{Goddard2012, SkitkaMosiere1999, MosierSkitka1996, KrepsTrustParadox}. This occurs as users attribute decision-making capabilities to the system beyond its intended design parameters and, in more severe cases, even favor its recommendations despite contradictory evidence \citep{KrepsTrustParadox}. Automation bias typically manifests in two forms: errors of omission, in which users fail to act because the automation issued no alert; and errors of commission, in which users act on incorrect or inappropriate guidance from the automation \citep{Goddard2012}.

Algorithm aversion is the phenomenon where individuals reject algorithmic or automated recommendations in favor of their own judgment \citep{AlgorithmicDecisionMaking, MoralDiscretion}. Algorithm aversion can occur even when users have evidence the algorithm would be more reliable than alternatives. Algorithm aversion can undermine the justified confidence users should have in AI given proven performance and transparency \citep{UNIDIRonCBM, ConferenceOnCBM}.

Automation bias and algorithm aversion are therefore core challenges to the effective, safe, and responsible use of AI in military and security contexts. Given that AI errors, at present, often manifest as plausible but false outputs rather than obvious failures, there is greater emphasis on automation bias than on aversion. This is reinforced by research on the psychology of error perception, which has consistently found that harmful inaction is perceived to warrant less blame and moral scrutiny than harmful actions, even when outcomes are equivalent\citep{SiuKit2022,Spranca1991,Vaal1996}. However, algorithm aversion can be equally as harmful in practice. When the USS Vincennes shot down Iran Air Flight 655 in 1988, the AEGIS Combat System correctly identified the aircraft as a civilian airliner, but the crew rejected this output, substituting an incorrect judgment under extreme stress.\citep{KahnCSET} This is also reflected in leading international normative efforts surrounding military AI, such as the \textit{Political Declaration on Responsible Military Use of Artificial Intelligence and Autonomy} \citep{PoliticalDeclaration}, endorsed by nearly 60 countries as of October 2024, and the United Nations Lethal Autonomous Weapons System Group of Governmental Experts (LAWS GGE) under the United Nations Convention on Certain Conventional Weapons (CCW) framework, both of which treat addressing automation bias as a central normative principle and an actionable safety measure. Provision G of the Political Declaration, for instance, urges that ``States should ensure that personnel who use or approve the use of military AI capabilities are trained so they sufficiently understand the capabilities and limitations of those systems in order to make appropriate context-informed judgments on the use of those systems and to mitigate the risk of automation bias,'' while the May 2025 provisional text of the LAWS GGE includes the directive to ``implement measures to detect, correct or mitigate, as much as possible, automation bias'' \citep{GGEonAWS}.

Despite understanding of the importance of automation bias and algorithm aversion, they remain under-theorized and insufficiently tested with regard to military AI DSS \citep{KahnCSET}. Studies of militaries have been limited, though existing work usefully demonstrates how some military sample are trusting of AI in theory but also have a "conservative understanding of the appropriate use and oversight of AI" \citep{lushenko2024artificial, lushenko2025ai}. States are still grappling with techniques and working to develop methodologies, policies, and technical practices that are most effective at mitigating automation bias. Some scholars warn that integrating AI DSS into militaries, if not done properly, risks shifting operators from active decision-makers to passive recipients of algorithmic output, leading to biased choices that could increase the risk of accidents and miscalculation at the tactical and operational levels \citep{AIDSSLegalRisks, bode2024problem, probasco2024not, garcia2024algorithms}. Doing so, several argue, could potentially erode appropriate human judgment in sensitive military tasks such as targeting, even undermining compliance with international humanitarian law \citep{IHLandAWS}. This concern is frequently projected onto the DSS as the technological artifact responsible, though it is more accurately a reflection of human error, unless human-designed standard operating procedures instruct operators to implement algorithmic recommendations without critical thinking \citep{AIDSSLegalRisks}. Previous research demonstrates that the quality of advice, specifically the accuracy rate of DSS, significantly influences automation bias. \citet{lyell_coiera_2016} reported that automation bias research consistently found that ``participants made significantly more AB [automation bias] errors when assisted by automation that was constantly highly accurate compared to automation that varied between high and low accuracy.'' Paradoxically, this suggests that higher algorithmic accuracy increases the likelihood of automation bias because automation bias is primarily a human factors and training issue. 


Given the increasing prevalence of AI DSS in military contexts and the potential consequences, understanding the extent to which military officers are susceptible to automation bias and algorithm aversion is important. The literature above suggests that military officers may be particularly subject to automation bias for several reasons. First, the need to accelerate decision-making on an increasingly automated battlefield could lead officers to delegate decision-making authority to AI DSS, even if such systems are known to introduce errors. This is relevant when countries are preparing for conventional combat operations, given that major powers such as the United States, China, and Russia are experimenting with integrating AI DSS into their command structures, thereby decreasing their sensor-to-shooter timelines \citep{AdlerMilitaryReview}. Second, military officers with an acute understanding of their own limitations and capabilities on an increasingly high-speed battlefield might over-rely on machine systems that promise greater accuracy, while ignoring signs of those systems' limits. 

Third, the average military officer may not have the training, knowledge, and context in AI necessary to understand AI risk at a technical level, thereby increasing the likelihood of automation bias, especially when using AI DSS. As a proxy for military officers, we rely on a sample of USMA cadets, which we justify in greater detail below. If military officers are more prone to automation bias, we should see the following:

\textit{H1: USMA cadets will exhibit higher levels of automation bias than a comparable general public sample.}

On the other hand, USMA cadets are AI natives; not only are they from a demographic group more likely to have significant experience with AI (though that would also be true of similarly situated members of the general public), but military training itself may make automation bias less likely. Unlike most colleges and universities, even those cadets not on an engineering track are introduced to military technologies, decision-making principles, and the notion of responsible use in general, to ensure they can effectively employ them in the military \citep{WestPointAOG}. Everyone from English majors to political science students becomes familiar with defense capabilities during their time at West Point, as they will all commission as second lieutenants in the Army after graduation. Given their set career trajectory, they have inherent incentives to ensure they can correctly use the types of technologies that are being deployed on the battlefield without falling prey to biases, as they can reasonably expect to have to use those technologies in life-or-death combat situations during the course of their careers.

Additionally, the USMA integrates AI into learning and dedicates effort to familiarizing cadets with the fundamentals of the technology and various applications \citep{USMAAI}. West Point ``provides cadets with a structured, progressive pathway from foundational concepts to applied expertise. All cadets develop core skills in human--machine teaming, partnering with AI and computing tools in responsible ways that enhance their ability to learn, solve problems, and think critically'' \citep{USMAAI}. For example, the 2024-2025 academic year theme was ``The Human and the Machine: Leadership on the Emerging Battlefield,'' designed to ``delve into the intricate relationship between human agency and technological innovation in contemporary warfare,'' exploring ``how leaders may effectively navigate the complexities of modern battlefields, where technological disruptions and advancements in autonomous systems, artificial intelligence, and cyber technologies are reshaping conflict dynamics'' \citep{USMATheme}. Existing research by Lushenko and Lushenko and Sparrow \citeyear{lushenko2024artificial, lushenko2025ai} described above also suggests West Point cadets will be optimistic about the possibilities of AI, but discerning in how they think about its use. If military officers are less prone to automation bias, we should see the following:

\textit{H2: USMA cadets will exhibit lower levels of automation bias than the general public.}

Unlike automation bias, where the direction of the effect is uncertain given existing research, algorithm aversion allows for a cleaner prediction. Whether cadets over-trust AI (H1) or are appropriately calibrated (H2), military training's emphasis on reliance on decision-support systems should suppress aversion in either case. In the case of H1, overtrust in AI precludes aversion toward it—here the two constructs do move inversely. But in the case of H2, appropriate calibration does not imply greater aversion; cadets should simply use AI as intended. This is because automation bias and algorithm aversion are conceptually distinct—high automation bias necessitates low algorithm aversion, but the absence of bias does not necessarily produce it. In all cases therefore, whether they are overconfident in AI or appropriately calibrated, cadets should have lower levels of algorithm aversion than the general public.

\textit{H3: USMA cadets will exhibit lower levels of algorithm aversion than a comparable general public sample.}

\section{Research Design}

To test the extent to which military officers are susceptible to automation bias when using AI DSS relative to the general public, we compare a scenario-based survey experiment conducted with West Point cadets to one conducted with members of the general public. The West Point experiment was fielded in May 2025. The cadets were all enrolled in SS307, the introductory international relations course at West Point; as a mandatory course corps-wide, this ensured that we surveyed a representative cross-section of USMA cadets, not limited by field of interest or study. The course director emailed cadets to request voluntary participation in the study.

Out of the 529 cadets contacted via email to participate in the study, 236 (44.6\%) completed the survey. To incentivize participation, the course director offered bonus points for survey completion, totaling 0.1\% of available points in the course. The survey took a median of 12 minutes to complete. Prior research shows that surveying West Point cadets can generate analytical leverage on questions surrounding military decision-making for several reasons \citep{jost2022character, brooks2022makes, dempsey2009our}. First, the U.S. Army overwhelmingly selects military leaders from West Point graduates, meaning they are more likely to represent future senior military leaders. Second, as explained above, the cadets receive AI-specific training. Third, all cadets go through Cadet Basic Training and other functional military training, differentiating them from the general public. The official West Point Military Program from the Department of Military Instruction describes the goal of the training as ``to instill in Cadets the foundational military competencies necessary for leaders to fight and win our Nation's wars,'' and all cadets take classes such as ``Introduction to Warfighting'' \citep[p.~8]{westpoint}.

An identical experiment was previously run on the general public, specifically \citet{horowitzkahnisq}'s representative sample of 9,000 adults from the United States, France, Australia, Japan, South Korea, Sweden, and the United Kingdom, as well as urban adults from China and Russia. We constructed a demographically matched sample of the general public for comparison with the West Point sample. Given that all West Point cadets surveyed were, by default, enrolled in university, we restricted the general public sample to respondents who reported having attended university. Additionally, given a maximum age of twenty-five for the West Point cadets, we likewise restricted the general public sample to individuals aged twenty-five or younger at the time of surveying, consistent with prior research showing that age is highly---and variably---correlated with openness to technologies and their use, particularly AI \citep{APYoungAdults, PewAISurvey, ZhangDafoe2019, localOfficialsAttitudes}. The resulting demographically aligned sample of the general public consisted of 702 respondents.

We contrast the general public and West Point cadets for three reasons. First, the comparison helps illustrate the extent to which concerns about automation bias in the general public are applicable to the military. Second, the comparison helps us start to identify the remediable and immediately addressable factors influencing automation bias for enhancing accurate use of AI systems, specifically AI DSS, in national security contexts among members of the armed services. Third, because we are comparing a subset of the general public with the West Point sample, the extent to which the general public might represent a proxy for elites is less important \citep{kertzer2022experiments, kertzer2022re, tetlock2009, TomzWeeks2013}.

The original survey was conducted in 2022, with the West Point survey conducted in 2025. Though three years apart, the surveys are appropriate for comparison for the following three reasons. First, the ability to respondents to complete the task does not change across the time period, because the airplane identification task, the core of the experiment, is static and not dependent on technological knowledge. Second, while LLMs are in the news and dominate AI conversations in elite populations, it's less clear there has been substantial change for the average person, especially outside the United States. Third, the most important different between the West Point and general public samples has to do with the training that the West Point cadets get, which does not skew the results of the general public sample from being older. Even if over time knowledge of AI has grown somewhat, the average person would still be in the early stages of knowledge from a Dunning-Kruger perspective, meaning we would not expect anything different today from a general public sample \citep{horowitzkahnisq}. After all, most people use AI today without realizing it \citep{maese}. Thus, there are not any systematic biases that mean we cannot compare the samples.

\subsection{Survey Design}

The experiments use a within-subject, pre--post design in which each participant's decisions were observed before and after exposure to experimental treatments. At the outset, respondents were instructed on the nature of the task: a pattern-recognition exercise that required making determinations of whether a given aircraft belonged to their country's military or to that of an adversary based on three specified identifying features (wing taper, wing tip shape, tail shape). An example of the exercise is visible in supplemental materials \ref{fig: Easy Difficulty Example Scenario} and \ref{fig: Hard Difficulty Example Scenario}. Respondents then completed five initial practice trials with no time pressure, no image obfuscation, and immediate feedback provided on the accuracy of the identification to ensure the task was understood and to establish a personalized baseline of task performance.

Participants then advanced to the experiment, where they completed a total of ten randomized aircraft classification scenarios. The first five were comparatively easier, followed by five of greater difficulty. Classification scenarios were defined by time constraints (either 10 or 7 seconds) and varying levels of feature obscuration (partial or full) to increase cognitive load, task complexity, and uncertainty. In each case (except for the more challenging levels in which one of the three distinguishing features was entirely obscured) respondents received sufficient visual information to allow for an accurate and verifiable identification.

In each scenario, respondents made an initial classification and were then presented with a decision aid---DSS---framed as a team member offering a recommendation. After receiving the recommendation, respondents were reminded of their initial identification and could either confirm their initial identification or revise it (see supplemental materials \ref{fig: Example Identification} for an example), except in the control condition, in which no post-identification recommendation was given. In these instances, respondents were advised that their team was unable to reach them and were not provided an opportunity to revise their identification.

The DSS was described either as an AI or a human analyst, with respondents told the extent of the prior testing and training of the system. High-confidence treatments were said to have ``undergone extensive testing and training to identify airplanes under these conditions,'' whereas low-confidence versions were described as ``still undergoing testing and being trained to identify airplanes under these conditions.'' We employed a two-by-four factorial structure with a control, yielding nine total conditions. The source of the recommendation (human vs. algorithm) and the confidence level (high vs. low) were crossed, and the correctness of the recommendation (friendly or enemy aircraft) was randomized. The nine possible treatment conditions were as follows:

\begin{enumerate}
        \item Human analyst low confidence (accurate)
        \item Human analyst high confidence (accurate)
        \item AI algorithm low confidence (accurate)
        \item AI algorithm high confidence (accurate)
        \item Human analyst low confidence (inaccurate)
        \item Human analyst high confidence (inaccurate)
        \item AI algorithm low confidence (inaccurate)
        \item AI algorithm high confidence (inaccurate)
        \item No identification suggestion
\end{enumerate}

After the ten scenario exercises, respondents subsequently answered a series of questions about their experience and familiarity with, knowledge of, and trust in AI and its applications, along with basic demographic information.

\subsubsection{Dependent Variables}

Unless otherwise noted, the dependent variable in the analyses below is a binary variable measuring whether a respondent ``switched'' their identification after interacting with the DSS (1 if post-treatment identification was different from the initial identification, 0 otherwise).

\subsubsection{Independent Variables}

The experiments use a standardized measure of a respondent's overall ``AI background,'' designed to capture varying dimensions and levels of knowledge, familiarity, and experience with AI in line with existing research on potential factors influencing automation bias \citep{horowitzkahnisq, Parasuraman2014, Schepman2020, Chong2022}. For the West Point sample, we collected demographic data alongside batteries of questions assessing respondents' exposure to, interactions with, and sentiments toward AI technologies. To operationalize these sub-indices, we relied on validated questions that capture distinct but related dimensions of respondents' relationship to AI, namely:

\textbf{AI Knowledge} -- Factual understanding of AI concepts and underlying technology, and measured using performance on two AI knowledge ``quiz'' questions.

\textbf{AI Familiarity} -- Exposure to and engagement with AI in public and interpersonal discourse; measured using a scale of how familiar respondents were with AI, based on whether they had heard about AI from friends or family, read about AI in the news, from other avenues, or not at all.

\textbf{AI Experience} -- Direct interaction with AI-based tools in practical day-to-day use and in formal educational or professional contexts. This subindex has three subcomponents:
    \begin{enumerate}
        \item whether the individual had some degree of coding or programming experience;
        \item whether the respondent had used specific applications and viewed them as employing AI; and
        \item whether they have an advanced degree in engineering/STEM.
    \end{enumerate}

The AI experience index differs slightly from the subindex generated in \citet{horowitzkahnisq}, which did not include degree type. Given that all respondents in the West Point sample had some university experience by default, and the general public sample was constructed to include those with university experience, we also included major/degree type as a contributing factor in the updated experience index.

\textbf{AI Background} -- An index aggregating knowledge, experience, and familiarity with AI. Knowledge, experience, and familiarity were each weighted equally ($\frac{1}{3}$ each). In the initial Horowitz and Kahn paper \citeyear{horowitzkahnisq}, experience and familiarity were each weighted at $\frac{2}{5}$, while knowledge was weighted at $\frac{1}{5}$ because the knowledge questions were difficult for respondents, with only 25 respondents out of 9,000 ($0.28\%$) answering them correctly in the original study \citep{horowitzkahnisq}. By contrast, among the West Point cadet sample, 33 respondents ($13.98\%$) answered both AI knowledge quiz questions correctly. Accordingly, we adjusted the scale to account for this increase.

All indices were normalized and (re)constructed identically for the West Point cadets and the demographically matched general-public sample, in line with the methodology described above. The specific survey questions and batteries used to comprise these indices are available in the supplemental materials.

\section{Results}

\subsection{Population Differences}

\begin{figure}
\centering
\includegraphics[width=0.9\columnwidth]{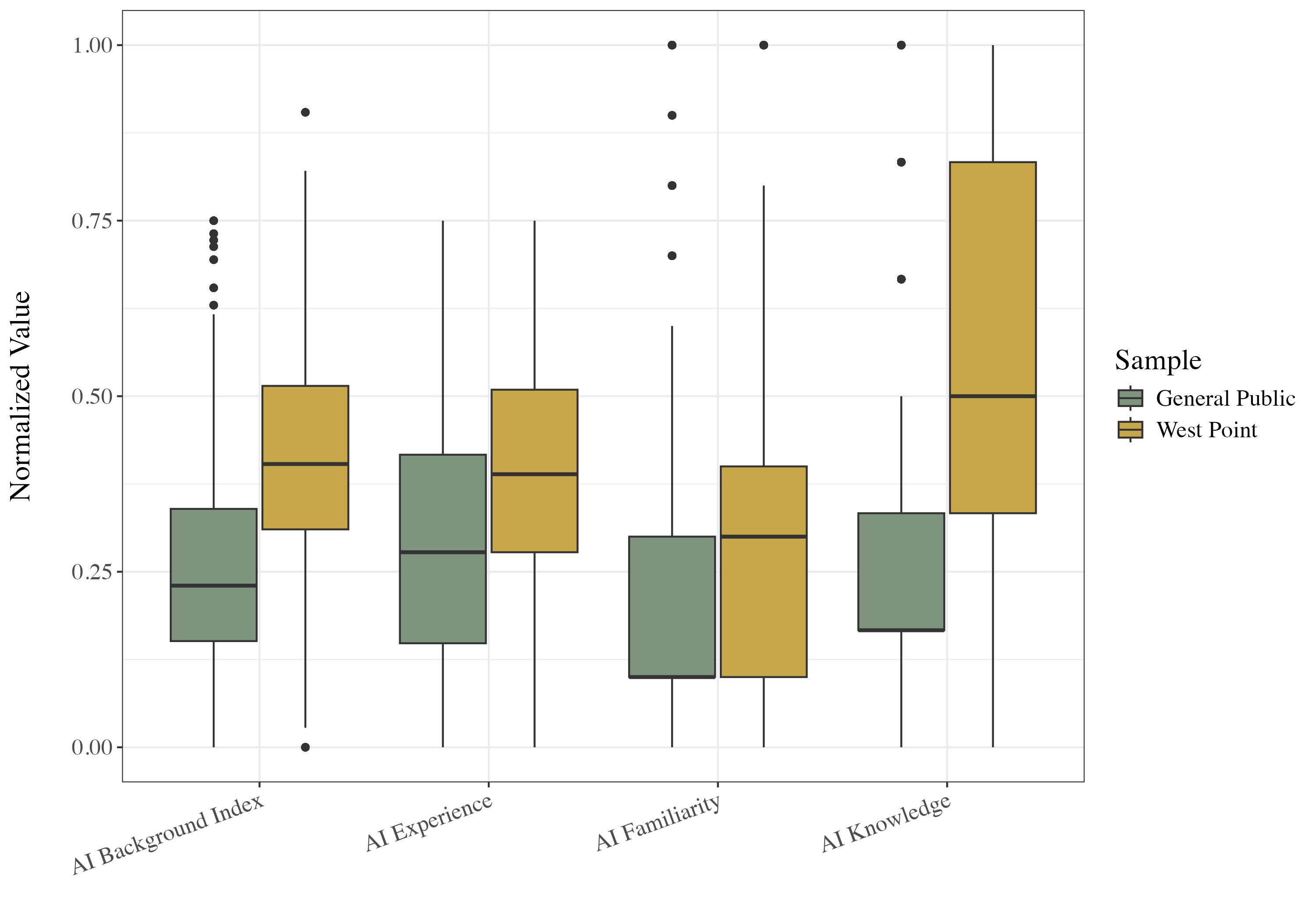}
\caption{AI Indicator Values, By Sample}
\label{fig: Box and Whiskers Plot}
\end{figure}

Tables \ref{tab: OG Sample Summary Stats} and \ref{tab: Westpoint Summary Stats} and Figure \ref{fig: Box and Whiskers Plot} below reports summary statistics for the general public sample and the West Point cadets (correlation matrices for each population sample are available in supplemental materials as tables \ref{fig: Correlation Matrix General Population} and \ref{fig: Correlation Matrix West Point}). As each respondent completed ten identification tasks, we report statistics at the task level rather than aggregating by respondent, thereby increasing the observation count by a factor of ten, though we statistically control for individual differences in our tests. The results are consistent whether we aggregate by respondent or not.

In both samples, the average age is nearly identical at 22 years, reflecting the restricted comparison set constructed for this study and validating the methodology employed to identify the general public sample. However, as expected given the institutional composition of the U.S. service academies, the West Point cadet sample shows a pronounced gender imbalance towards males (72\%), compared with the general-public sample (43\%).

\begin{center}
\begin{table}[!htbp] \centering 
  \caption{Descriptive Statistics (Comparable General Public)} 
  \label{tab: OG Sample Summary Stats}
\begin{tabular}{@{\extracolsep{5pt}}lccccc} 
\\[-1.8ex]\hline \\[-1.8ex] 
Statistic & \multicolumn{1}{c}{N} & \multicolumn{1}{c}{Mean} & \multicolumn{1}{c}{St. Dev.} & \multicolumn{1}{c}{Min} & \multicolumn{1}{c}{Max} \\ 
\hline \\[-1.8ex] 
Sex (Female = 1) & 6,980 & 0.57 & 0.49 & 0 & 1 \\ 
Age & 7,020 & 22.07 & 2.00 & 18 & 25 \\ 
Education in Stem & 7,020 & 0.36 & 0.48 & 0 & 1 \\ 
Practice Round Accuracy & 7,020 & 0.52 & 0.25 & 0.00 & 1.00 \\ 
Received a High Confidence Treatment & 7,020 & 0.45 & 0.50 & 0 & 1 \\ 
Received a Low Confidence Treatment & 7,020 & 0.44 & 0.50 & 0 & 1 \\ 
Received an AI Algorithm Treatment & 7,020 & 0.45 & 0.50 & 0 & 1 \\ 
Received a Human Analyst Treatment & 7,020 & 0.44 & 0.50 & 0 & 1 \\ 
Switched Identification After Treatment & 6,254 & 0.23 & 0.42 & 0 & 1 \\ 
AI Background Index & 7,020 & 0.25 & 0.14 & 0.00 & 0.75 \\ 
Experience in AI Index & 7,020 & 0.29 & 0.17 & 0.00 & 0.75 \\ 
Perceived AI Usage & 7,020 & 0.43 & 0.29 & 0.00 & 1.00 \\ 
Programming Experience & 7,020 & 0.35 & 0.34 & 0.00 & 1.00 \\ 
Familiarity in AI Index & 7,020 & 0.20 & 0.18 & 0.00 & 1.00 \\ 
Knowledge in AI Index & 7,020 & 0.27 & 0.21 & 0.00 & 1.00 \\ 
Trust in AI & 6,010 & 0.59 & 0.13 & 0.11 & 0.93 \\ 
\hline \\[-1.8ex] 
\end{tabular} 
\end{table} 
\end{center}

\begin{center}
\begin{table}[!htbp] \centering 
  \caption{Descriptive Statistics (West Point)} 
  \label{tab: Westpoint Summary Stats}
\begin{tabular}{@{\extracolsep{5pt}}lccccc} 
\\[-1.8ex]\hline \\[-1.8ex] 
Statistic & \multicolumn{1}{c}{N} & \multicolumn{1}{c}{Mean} & \multicolumn{1}{c}{St. Dev.} & \multicolumn{1}{c}{Min} & \multicolumn{1}{c}{Max} \\ 
\hline \\[-1.8ex] 
Sex (Female = 1) & 2,350 & 0.28 & 0.45 & 0 & 1 \\ 
Age & 2,360 & 22.77 & 0.96 & 21 & 26 \\ 
Education in Stem & 2,360 & 0.59 & 0.49 & 0 & 1 \\ 
Practice Round Accuracy & 2,360 & 0.77 & 0.23 & 0.00 & 1.00 \\ 
Received a High Confidence Treatment & 2,360 & 0.44 & 0.50 & 0 & 1 \\ 
Received a Low Confidence Treatment & 2,360 & 0.44 & 0.50 & 0 & 1 \\ 
Received an AI Algorithm Treatment & 2,360 & 0.44 & 0.50 & 0 & 1 \\ 
Received a Human Analyst Treatment & 2,360 & 0.44 & 0.50 & 0 & 1 \\ 
Switched Identification After Treatment & 2,072 & 0.14 & 0.35 & 0 & 1 \\ 
AI Background Index & 2,360 & 0.41 & 0.15 & 0.00 & 0.90 \\ 
Experience in AI Index & 2,360 & 0.38 & 0.16 & 0.00 & 0.75 \\ 
Perceived AI Usage & 2,360 & 0.58 & 0.28 & 0.00 & 1.00 \\ 
Programming Experience & 2,360 & 0.42 & 0.30 & 0.00 & 1.00 \\ 
Familiarity in AI Index & 2,360 & 0.32 & 0.21 & 0.00 & 1.00 \\ 
Knowledge in AI Index & 2,360 & 0.53 & 0.28 & 0.00 & 1.00 \\ 
Trust in AI & 2,210 & 0.61 & 0.14 & 0.18 & 0.93 \\ 
\hline \\[-1.8ex] 
\end{tabular} 
\end{table} 
\end{center}

\subsection{Comparing Views of Artificial Intelligence}

We begin by assessing background and trust in AI among West Point cadets and the comparable general public sample. Across each index, West Point cadets scored statistically significantly higher ($p<0.001$ for every comparison, see \ref{tab: T-Test Table} in the supplemental materials).

The cadets' mean AI Background Index is $0.41$ compared to $0.25$ for the comparable general public sample, and their AI Knowledge score is nearly twice as high. Cadets likewise have greater levels of experience and familiarity with AI and are substantially more likely to possess a STEM-related educational background.

Figure \ref{fig: Box and Whiskers Plot} further illustrates these differences, with the West Point sample exhibiting higher median and, often, much higher minimum values across each AI-related index. The cadets thus entered the experiment with systematically stronger AI-related backgrounds. Some of the variation and increase across all vectors in the 2025 West Point cadet sample is undoubtedly due to the greater exposure to AI tools since the fielding of the original survey experiment in 2022, though the comparison is valid for the reasons explained above. 

Crucially, in the context of \citet{horowitzkahnisq}'s findings that automation bias follows a Dunning-Krueger-like pattern across AI background experience levels \citep{SanchezDunning2018}, the West Point population starts further along the curve towards accurate calibration than the comparable general public sample. These findings even before assessing the experimental results suggest some initial support for hypothesis 2 and hypothesis 3, given the way greater knowledge and education about AI should decrease the potential for automation bias and for algorithm aversion.

In addition to respondents' overall knowledge, familiarity, and experience with AI, we also measured their trust in AI \citep{Chong2022, AquilinoAIReview, DramschAITrust, OrbanAITrust}, using a reconstructed version of the AI trust battery developed by Horowitz and Kahn $($\citeyear{horowitzkahnisq}$)$. Measuring trust helps identify baseline susceptibility to automation bias.

\begin{figure}
    \centering
    \includegraphics[trim={0 1cm 0 0},clip,width=\columnwidth]{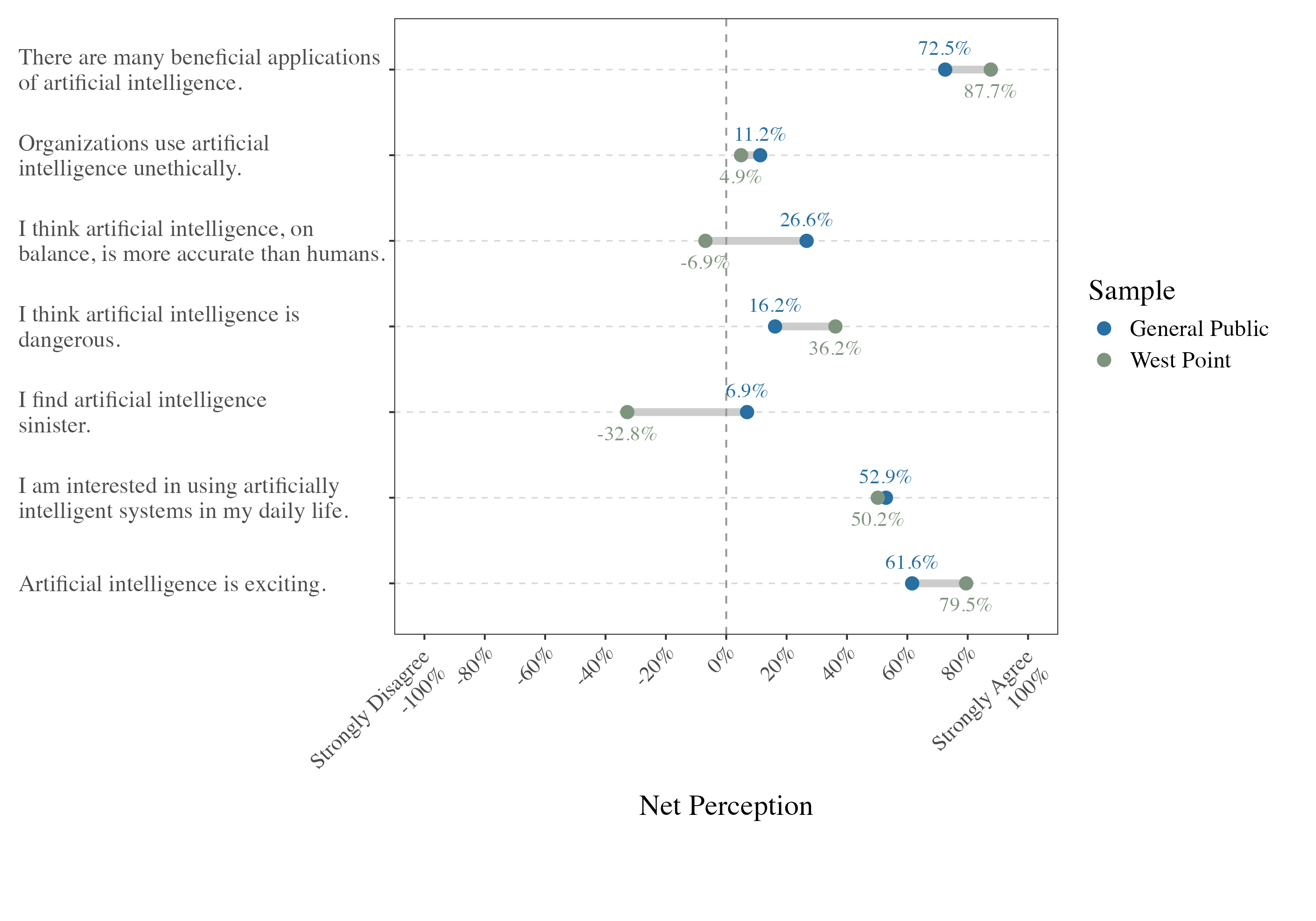}
    \caption{Views of AI, By Sample}
    \label{fig: Net Perception Dumbbell}
\end{figure}

The trust battery consists of seven statements that assess respondents' trust in artificial intelligence and AI-enabled systems. The statements are drawn from the top three factor-loaded items from both the positive and negative subscales of an AI-specific attitudinal index adapted from the Technology Readiness Index 2.0, a validated framework for measuring perceptions of new technologies \citep{Parasuraman2014, Lam2008, Schepman2020}. Figure \ref{fig: Net Perception Dumbbell} shows the differences in AI net perceptions for each population sample across these seven statements.

Net perception is the percentage of respondents who agree minus the percentage who disagree, with values ranging from $-100$ (uniform disagreement) to $+100$ (uniform agreement) (e.g., if $70\%$ disagree, and $50\%$ agree, net perception is $-20\%$). Further granularity, including the breakdown of each sample's responses to these statements, is available in figures \ref{fig: Likert General Population} and \ref{fig: Likert West Point}. Overall, both West Point cadets and the general public are open to using AI: each group shows relatively strong interest in using AI applications in daily life, with net agreement just over $50\%$. West Point cadets are especially positive about the potential benefits of AI. They report even higher net agreement than the general public that there are many beneficial applications of artificial intelligence ($87.7\%$ compared to $72.5\%$), and are more likely to find AI exciting ($79.5\%$ versus $61.6\%$).

However, these relatively positive views of West Point cadets are tempered by concerns about AI's limitations. West Point cadets also express more caution than the general public about AI, particularly regarding AI's accuracy and the potential dangers associated with its use. For example, cadets are much less likely to describe AI as ``sinister'' ($-32.8\%$ net disagreement, versus the public's $6.9\%$ net agreement), and they are considerably more likely to worry about the dangers of artificial intelligence ($36.2\%$ of cadets versus $16.2\%$ of the public). Cadets' skepticism appears grounded in concerns about AI's real-world consequences, and whether it can reliably outperform humans, rather than by attributing malice to the technology itself. This is reflected in their net disagreement ($-6.9\%$) that AI is more accurate than humans, a sharp contrast to the general public's clear net agreement ($26.6\%$). While both populations are receptive to AI's promise, West Point cadets maintain a distinctly more cautious perspective than their non-military peers. This provides more support for hypothesis 2.

While the causal mechanism is beyond the scope of this paper, it is possible that West Point cadets may be following best practices for addressing automation bias, given the training they receive, making them more accurately calibrated about AI use \citep{kupfer2023check, KahnCSET}. Research on automation bias shows that the most effective design tactics for mitigating automation bias are those that actively promote critical engagement with, and independent verification of, algorithmic outputs \citep{RomeoandConti}. In contrast, simplistic and cognitively demanding explanations can make misplaced trust more likely, especially among groups with low AI background levels.

\subsection{Switching Judgments In Response to Decision Support Systems}

\begin{figure}
\centering
\includegraphics[width=0.9\columnwidth]{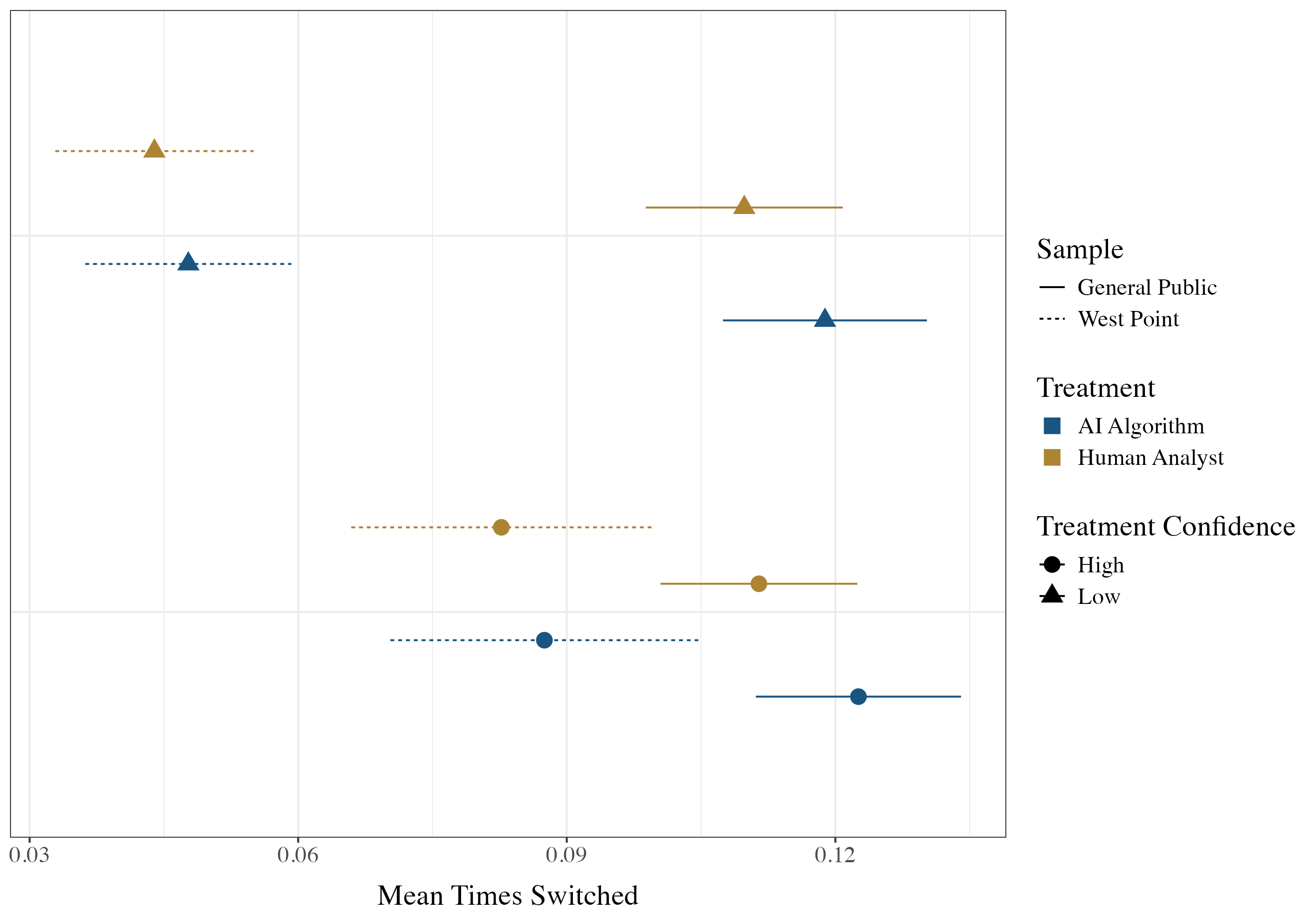}
\caption{Switching Rate Across Samples and Experimental Conditions}
\label{fig: Rate of Switching Plot}
\end{figure}

We now assess differences between the samples using the primary dependent variable: rate of switching. Figure \ref{fig: Rate of Switching Plot} presents the mean switching rate for respondents across the samples and treatment conditions without any control variables.

Across the control and treatment conditions in both the general public and West Point samples, respondents were most likely to change their original identifications when the DSS was described as an AI algorithm, and more so in the high-confidence condition. Conversely, the low-confidence human analyst treatment consistently yielded the lowest response-switching frequency, suggesting an overall bias towards automation. Switching occurred at consistently higher rates among the general public sample across all treatment conditions. West Point respondents exhibited statistically significant (at the $p <0.001$ level) lower baseline switching behavior at nearly half the rate (0.065 vs. 0.116), suggesting greater reluctance to revise decisions regardless of the treatment.

Overall, differences within the general public subsample across treatment conditions were not statistically significant, showing relative insensitivity to both the AI-versus-human distinction and confidence-level manipulations. For the broader public sample which includes all age and education categories, those differences are statistically significant, with switching especially more likely in the high-confidence treatment condition \citep{horowitzkahnisq}. Descriptively, for the sample, mean switching rates were slightly higher for AI conditions ($0.121$) than human conditions ($0.111$), while high- and low-confidence conditions showed virtually no difference ($0.117$ vs. $0.114$).

In contrast, while the West Point cadets exhibited much lower overall switching rates than the general public, they showed clear sensitivity to the confidence treatment rather than to whether systems were AI- or human-based. Specifically, cadets switched nearly twice as often in high-confidence conditions ($0.085$) compared to low-confidence conditions ($0.046$), a statistically significant difference ($p < 0.001$). In contrast, the difference between AI conditions ($0.065$) and human conditions ($0.061$) was marginal and not statistically significant. These findings align with the way the USMA seeks to train cadets so they can achieve justified confidence, that is, to be properly calibrated so their expectations of an AI system's accuracy match the reality of the accuracy of the system \citep{USMAAI, AIandLSCO}.

Taken together, the switching patterns and the trust battery responses show that West Point cadets more strongly respond to signals of justified confidence, such as the degree of training, error rates, or the extent of test, evaluation, verification, and validation (TEVV), rather than just the type of decision-support system, than relatively similar members of the general public. This provides additional support for hypothesis 2 and hypothesis 3.

\subsection{Evaluating Task Accuracy Across Samples}

Switching behavior and the trust battery provide insight into general trends regarding openness to AI, the approaches of these two samples to using AI, and how they might affect the prevalence of automation bias, but they do not capture the actual consequences that result from the occurrence of automation bias. To evaluate how reliance on decision aids affected actual outcomes, we now examine task accuracy. The analysis above is agnostic to the accuracy of the treatment suggestion due to the randomized experimental conditions, some of which made it impossible for respondents to accurately determine the right answer. We now focus on respondent accuracy at the airplane identification task.

To test for automation bias, we filtered for instances in which participants received an AI treatment that contradicted their initial identification and subsequently switched their identification to align with the recommendation of the treatment. The analysis was limited to rounds of ``easier'' difficulty in which respondents could visually confirm the correct answer by pattern matching against the provided guide. In more difficult rounds, in which one feature was entirely obscured, independent verification was not possible, meaning a respondent could not verify visually whether the DSS provided them with the correct advice. Outcomes were coded along two dimensions: whether the treatment DSS was correct or incorrect in its recommendation, and whether the respondent followed the DSS recommendation (defined as switching their answer to align with the treatment), resulting in the following possible outcomes:

\begin{itemize}
    \item Following incorrect advice was coded as ``automation bias.''
    \item Rejecting correct advice was coded as ``algorithm aversion.''
    \item Rejecting incorrect advice was coded as ``correct reject.''
    \item Following correct advice was coded as ``correct use.''
\end{itemize}

\begin{figure}[H]
\centering
\includegraphics[trim={0 7cm 0 7cm},clip, width=\columnwidth]{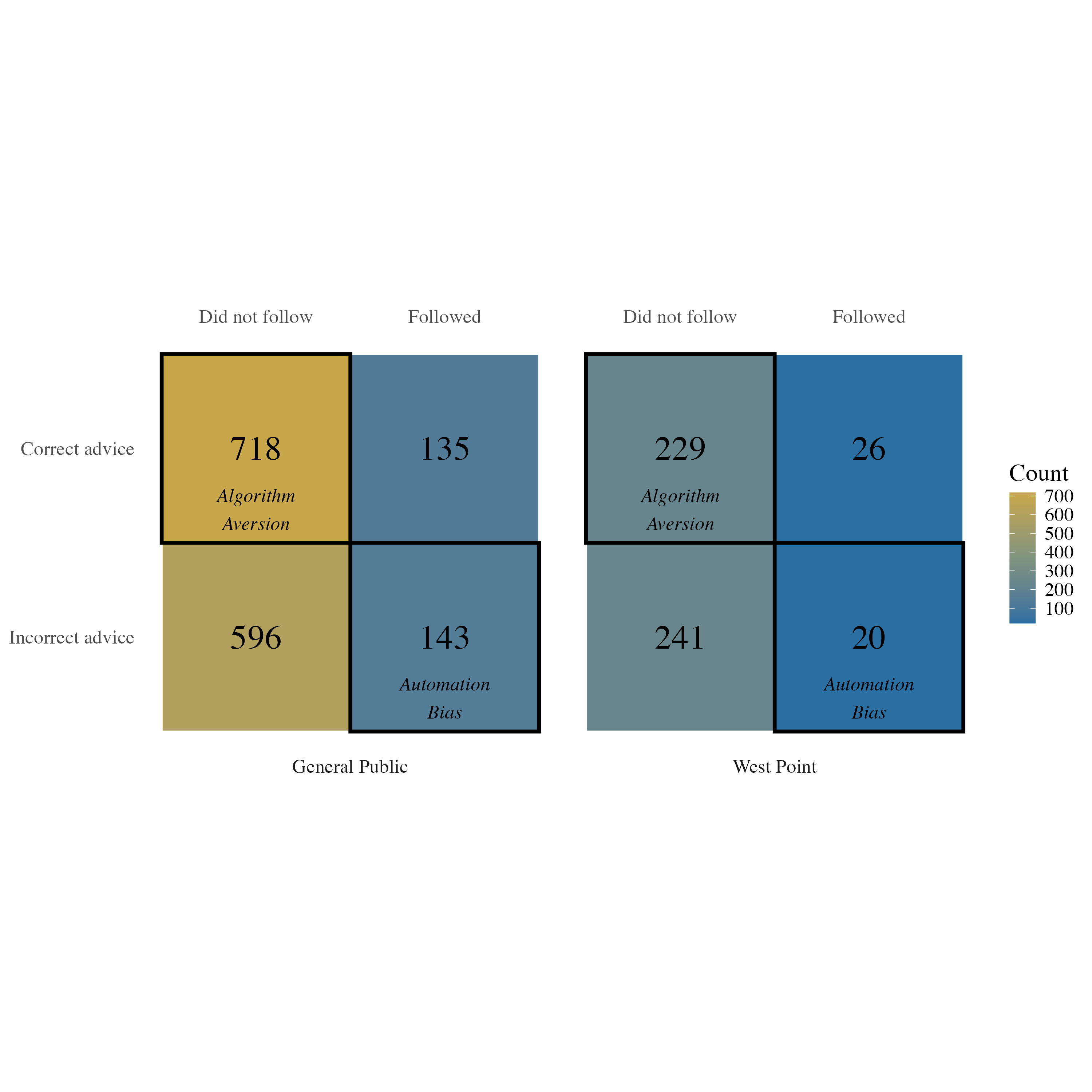}
\caption{Automation Bias and Algorithm Aversion, By Sample}
\label{fig: Automation Bias Rates}
\end{figure}

The resulting counts for each outcome are presented in the 2x2 table for each population sample in Figure \ref{fig: Automation Bias Rates}.

We further compare the differences across samples in Table \ref{tab: Error Rates}. The general public sample ($N = 1,592$) followed incorrect advice in $9\%$ of trials, whereas West Point cadets ($N = 516$) did so in only $3.9\%$ of trials. A two-sample test for equality of proportions with a continuity correction confirmed a statistically significant difference ($\chi^2=13.54, p < 0.001$), with West Point cadets demonstrating a rate of automation bias that was less than half that observed in the general public. In contrast, algorithm aversion rates were similar across samples and did not differ significantly (general public: $45.1\%$; West Point: $44.4\%$).

Beyond these between-sample differences, within-sample McNemar tests revealed that algorithm aversion was significantly more common than automation bias in both populations, demonstrating that participants were more likely to discount correct advice than to accept incorrect advice in general. This is evidence against hypothesis 3 because it shows statistically similar levels of algorithm aversion across populations. This asymmetry is nonetheless substantively notable as it suggests that whatever training and experience leads West Point cadets to be less susceptible to automation bias does not appear to similarly reduce their reluctance to follow correct advice.

One limitation of the experimental design is that it focuses on only one manifestation of automation bias: errors of commission, or the tendency to follow incorrect information presented by an AI DSS. Omission errors, such as failing to act when not prompted by a system, could not be assessed because respondents were not provided an opportunity to correct information without first being presented with a treatment. In the control condition, where no treatment was provided, respondents only made an initial identification before moving on to the next round. A related constraint applies to the measurement of algorithm aversion, in that in this study we operationalize it narrowly as the failure to follow correct algorithmic advice. In this context, it operates as a behavioral proxy that captures one dimension of aversion but cannot fully distinguish active distrust of the system from other reasons for non-switching, such as general confidence in one's initial answer. Future research could employ designs that examine omission-type automation bias, or more directly elicit algorithm aversion, for instance by allowing respondents to actively override or dismiss recommendations, or by measuring trust in the system before and after treatment exposure.

\begin{center}
\begin{table}[!htbp]
\centering
\caption{Automation Bias and Algorithm Aversion by Sample}
\label{tab: Error Rates}
\begin{tabular}{p{10cm}|c|c}

\hline
Statistic & General Public & West Point \\
\hline

$N$ 
& 1,592& 516\\

Automation Bias (\%)& 9.0& 3.9\\

Automation Bias (Count)& 143& 20\\

Algorithm Aversion (\%)& 45.1& 44.4\\

Algorithm Aversion (Count)& 718& 229\\

Automation Bias vs. Algorithm Aversion \newline(within-sample $\chi^2$)& 683.08***& 339.03***\\

Automation Bias: West Point vs. General Public (between-sample $\chi^2$)& \multicolumn{2}{c}{13.54***}\\

Algorithm Aversion: West Point vs. General Public (between-sample $\chi^2$)& \multicolumn{2}{c}{0.06}\\

\hline
\end{tabular}

\vspace{0.5em}
\footnotesize
\textit{Note:} McNemar’s $\chi^2$ tests evaluate within-sample asymmetry between automation bias and algorithm aversion. Between-sample $\chi^2$ tests of proportions compare the prevalence of each outcome between the General Public and West Point samples. Stars denote statistical significance: *** $p < 0.001$, ** $p < 0.01$, * $p < 0.05$.
\end{table}

\end{center}

\subsection{Regression Analysis}

To further test our hypotheses, we turn to regression analysis to evaluate differences between the West Point and general public samples while accounting for similarities in AI background index, trust in AI, performance on the initial five practice rounds (during which respondents received feedback on their performance accuracy), sex, and age.

To ensure that our analysis focused specifically on instances where automation bias could occur, we again restricted the test to experimental trials in which respondents received incorrect advice from the AI DSS at the easy difficulty level. Automation bias was measured as a binary dependent variable indicating whether the participant followed the incorrect AI recommendation on that identification task. This approach isolates cases in which participants had the opportunity to follow erroneous advice, providing a clearer test of susceptibility to automation bias. Table \ref{tab: Regression GLM} below shows the results. As each respondent completed multiple identification tasks, we report task-level statistics while controlling for individual differences using clustered standard errors at the participant level.

\begin{center}
\begin{table}[ht]
\centering
\caption{Logit Regression Predicting Automation Bias}
\label{tab: Regression GLM}

\begin{tabular}{l c}
\toprule
 & B/SE \\
\midrule
West Point (vs. General Public) & -0.937* (0.371) \\
AI Beliefs & 0.071 (0.116) \\
AI Background & 0.069 (0.116) \\
Practice Percentage Correct & -0.067 (0.120) \\
Age & 0.120 (0.107) \\
Sex (Female = 1) & 0.071 (0.201) \\
\midrule
Number of Observations & 870 \\
AIC & 743.5 \\
BIC & 776.9 \\
RMSE & 0.36 \\
Log Likelihood & -364.750\\
Pseudo R² & 0.024\\
\bottomrule
\multicolumn{2}{l}{\footnotesize Note: Standard errors in parentheses, clustered at the participant level.  ** $p<0.001$, ** $p<0.01$, * $p<0.05$.}\\
\end{tabular}
\end{table}

\end{center}

The results from a logit model with clustered standard errors by respondent show that participants from West Point had a statistically significantly lower likelihood of making errors of commission than the general public, consistent with our above findings. None of the other predictors were statistically significant in either a logit or a linear regression model (See \ref{tab: Regression OLS} in the Appendix).

Figure \ref{fig: Predicted Probabilities Automation Bias} shows the covariate-adjusted predicted probabilities of automation bias errors by sample in the logit model, with $95\%$ confidence intervals accounting for clustering at the participant level. As expected given the direction and significance of the statistical results, the West Point cadets surveyed consistently exhibit lower predicted probabilities (mean: $8.3\%$, $95\%$ confidence interval: $[4.4\%,12.1\%]$) of automation bias than general public participants (mean: $18.2\%$, $95\%$ confidence interval: $[15.0\%,21.3\%]$).

\begin{figure}[H]
\centering
\includegraphics[width=0.7\columnwidth]{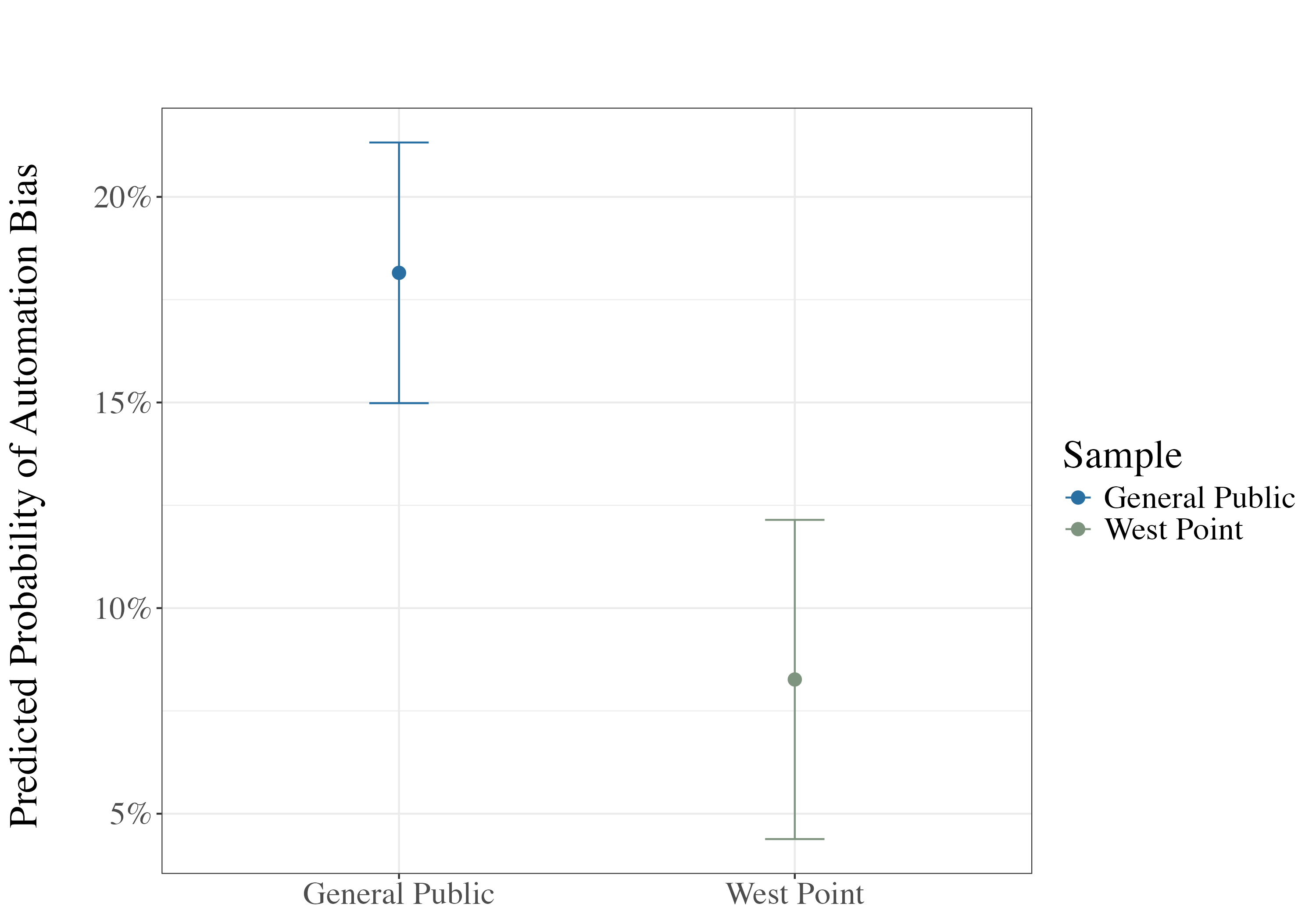}
\caption{Predicted Probabilities of Automation Bias}
\label{fig: Predicted Probabilities Automation Bias}
\end{figure}

\section{Conclusion}

How humans engage with AI systems to make decisions with national security consequences is a critical question for political science. Some existing research points to automation bias and algorithm aversion as mechanisms that could substantially increase the potential for accidents, miscalculation, and even inadvertent escalation as AI integration in militaries expands. By comparing a sample of West Point cadets with a similar general public sample, we demonstrate that the cadets, future U.S. military leaders, are better calibrated in their views of AI and less susceptible to automation bias. In contrast to some existing literature, our results provide the first direct experimental evidence on automation bias in a military population, and suggest that the desire of militaries to adopt AI will not necessarily lead to the use of flawed AI systems and could even offer opportunities for improvement

The results provide some initial evidence that, with training and habituation, those responsible for using AI in the military domain can apply it in a safe and effective way. No crisis or combat situation is free from errors in judgment, but militaries with training in decision-making in general, and AI-specific training, can decrease the risk of automation bias. The results also show that achieving justified confidence requires addressing automation bias and algorithm aversion through potentially distinct mechanisms, and that reducing one does not automatically diminish the other. Future research should examine these phenomena separately to better understand what interventions can move operators toward appropriate calibration along both dimensions.

The results here do not suggest that there is nothing to fear from AI DSS from an international relations perspective. The risks of military adoption of AI DSS do not solely arise from automation bias. If standard operating procedures, for example, instruct military personnel to trust AI algorithms more than their accuracy warrants, it could lead to adverse results as well. The casual logic in that case would be attributable to upper-echelon military organizational decisions, rather than to an individual operator exhibiting automation bias. This is potentially testable in future research through team-based laboratory experiments.

Fruitful avenues for future research include conducting similar experiments with non-commissioned officers, officers from different branches of the military, and mid- to late-career officers---all groups who may be more (or less) prone to automation bias and algorithm aversion due to age, life experience, and differences in training and education. If future surveys find that other groups within the military are more prone to cognitive distortions when using AI DSS, then the West Point educational and training model might suggest lessons learned for training for other groups.

Finally, this paper can also shape future research on crisis decision-making. While much of the literature assumes that cognitive biases are innate and difficult to change, the results here suggests that educational interventions and training can mitigate certain cognitive biases associated with AI DSS use, meaning potentially not all is as hopeless as the classic literature on crisis decision-making might suggest.

\bibliographystyle{apalike}
\bibliography{paper.bib}

\newpage

\appendix

\pagenumbering{arabic}

\counterwithin{figure}{section}
\counterwithin{table}{section}

\section{Online Appendix for What is Human in Judgment? Testing Automation Bias and Algorithm Aversion Among United States Military Academy Cadets}

\textbf{Trust in AI Battery}
These questions were used to generate the ``Trust in AI'' measure in the paper.

Please indicate the extent to which you agree or disagree with each of the following statements:
\begin{itemize}
    \item ``Organizations use artificial intelligence unethically.''
    \item ``I am interested in using artificially intelligent systems in my daily life.''
    \item ``I find artificial intelligence sinister.''
    \item ``I think artificial intelligence is dangerous.''
    \item ``Artificial intelligence is exciting.''
    \item ``I think artificial intelligence, on balance, is more accurate than humans.''
    \item ``There are many beneficial applications of artificial intelligence.''
\end{itemize}

Answer options:
\begin{itemize}
    \item Strongly disagree
    \item Somewhat disagree
    \item Neither agree nor disagree
    \item Somewhat agree
    \item Strongly agree
    \item I don't know
\end{itemize}

\textbf{AI Familiarity Battery}

What kind of exposure have you had to the topic of artificial intelligence (AI)? (Select all that apply)
\begin{itemize}
    \item I have heard about AI in the news, from friends or family, etc.
    \item I closely follow AI related news
    \item I have some formal education or work experience relating to AI
    \item I have extensive experience in AI research and/or development
    \item I have never heard of AI
    \item I don't know
\end{itemize}

\textbf{AI Experience Battery}

Do you have programming experience?
\begin{itemize}
    \item Yes, a great deal
    \item Yes, some
    \item No
    \item I don't know
\end{itemize}

Have you used any of the following?
\begin{itemize}
    \item A website or application that translates languages (e.g. Google Translate, Microsoft Translator)
    \item An application that identifies or categorizes people or objects in your photos or videos
    \item A self-driving car
    \item A chatbot that offers advice or customer support
    \item A digital personal assistant on your phone or a device in your home that can help schedule meetings, answer questions, and complete tasks (e.g. Alexa, Siri, Cortana)
    \item A robot, such as those used in work contexts or which act as social assistants
    \item A website or application that recommends movies or television shows based on your prior viewing habits
    \item A website that suggests advertisements for you based on your browser history
    \item A search engine (e.g. Google or Microsoft Bing)
\end{itemize}

Answer options:
\begin{itemize}
    \item Yes
    \item No
    \item I don't know
\end{itemize}

As far as you know, do each of the following applications use AI?
\begin{itemize}
    \item A website or application that translates languages (e.g. Google Translate, Microsoft Translator)
    \item An application that identifies or categorizes people or objects in your photos or videos
    \item A self-driving car
    \item A chatbot that offers advice or customer support
    \item A digital personal assistant on your phone or a device in your home that can help schedule meetings, answer questions, and complete tasks (e.g. Alexa, Siri, Cortana)
    \item A robot, such as those used in work contexts or which act as social assistants
    \item A website or application that recommends movies or television shows based on your prior viewing habits
    \item A website that suggests advertisements for you based on your browser history
    \item A search engine (e.g. Google or Microsoft Bing)
\end{itemize}

Answer options:
\begin{itemize}
    \item Yes
    \item No
    \item I don't know
\end{itemize}

\textbf{AI Knowledge Battery}

Choose the option(s) that is correct regarding artificial intelligence (AI). Select all that apply.
\begin{itemize}
    \item Artificial intelligence includes techniques that allow systems to learn without being explicitly programmed
    \item Artificial intelligence is a software, machine, or computer that researchers think could eventually emulate the human mind
    \item Machine learning is a type of artificial intelligence
    \item None of the above
    \item I don't know
\end{itemize}

What types of Artificial Intelligence are there?
\begin{itemize}
    \item Supervised, Unsupervised and Reinforcement Learning
    \item Monitored, Unsupervised and Signal Learning
    \item Signal Learning, Concept Learning and Rule Learning
    \item Reinforcement Learning, Unsupervised Learning and Signal Learning
    \item None of the above
    \item I don't know
\end{itemize}

\begin{figure}[H]
    \centering
    \includegraphics[width=\columnwidth]{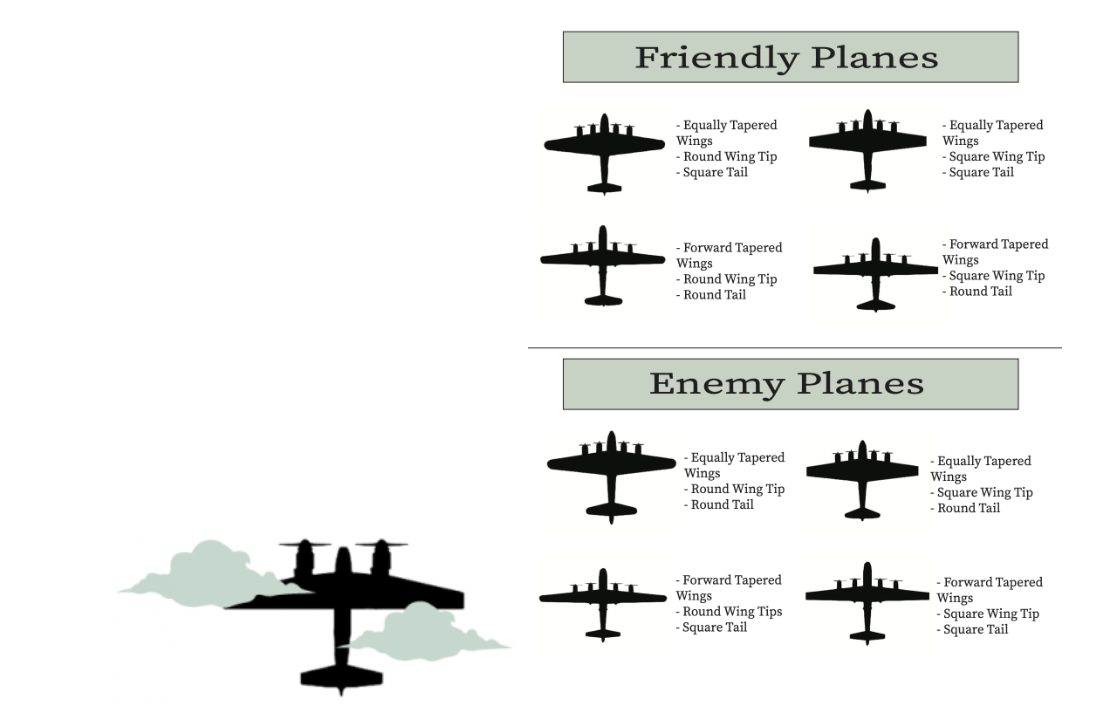}
    \caption{Example Identification Task - Easy}
    \label{fig: Easy Difficulty Example Scenario}
\end{figure}

\begin{figure}
    \centering
    \includegraphics[width=\columnwidth]{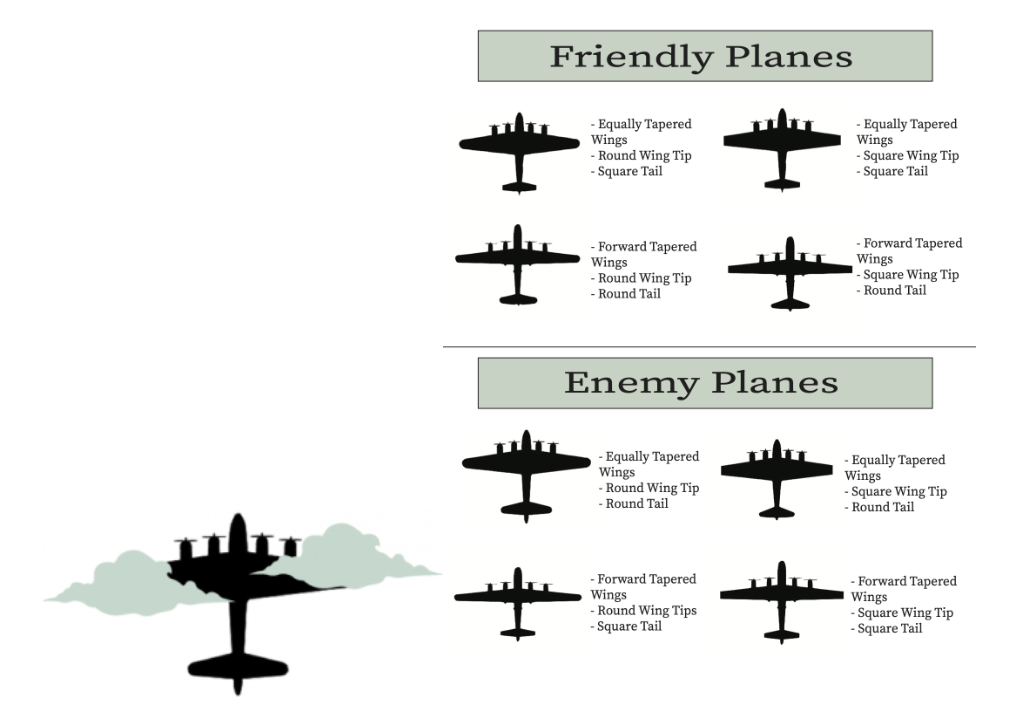}
    \caption{Example Identification Task - Difficult}
    \label{fig: Hard Difficulty Example Scenario}
\end{figure}

\begin{figure}
    \centering
    \includegraphics[width=\columnwidth]{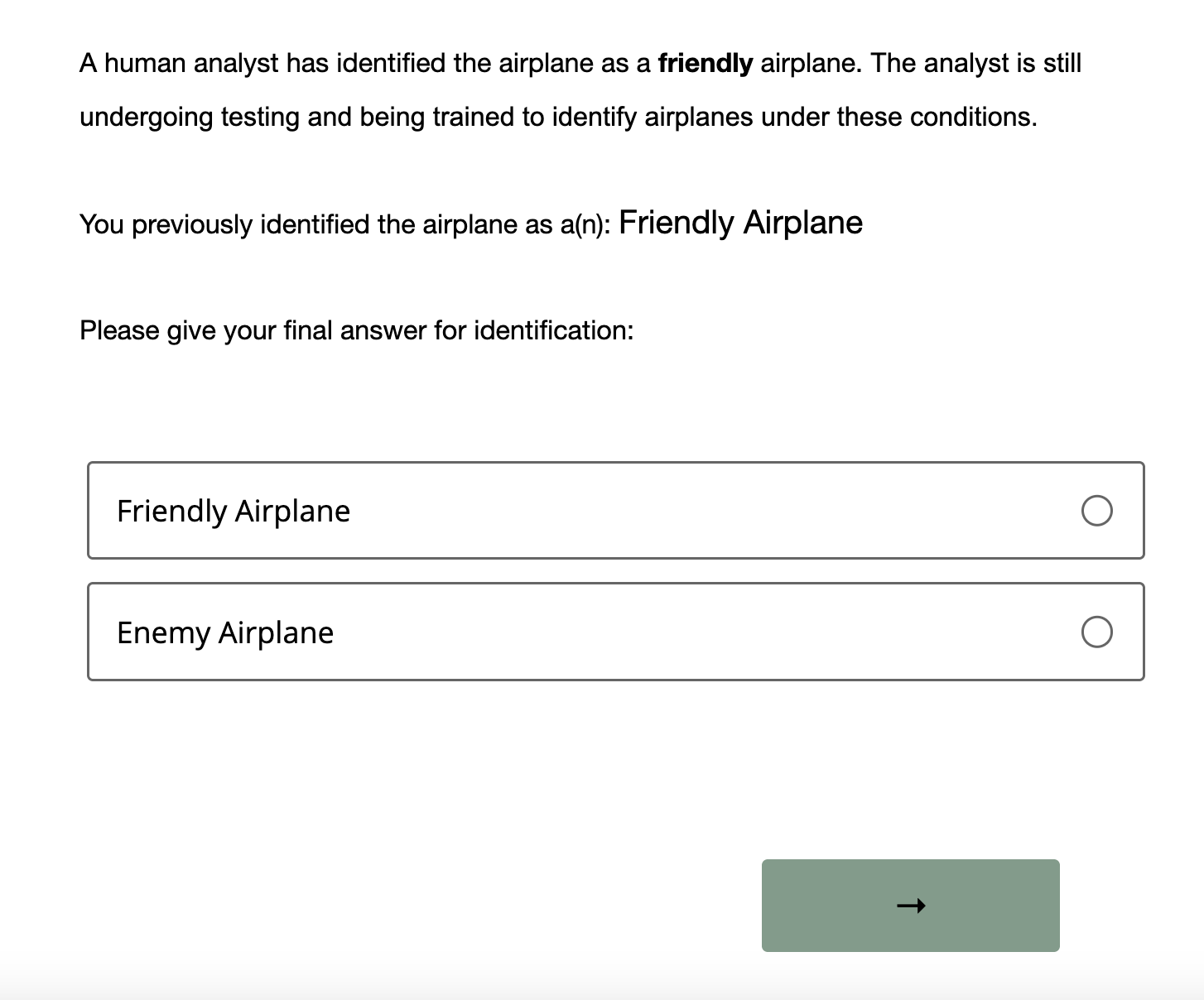}
    \caption{Example Treatment Scenario}
    \label{fig: Example Identification}
\end{figure}

\begin{center}
    \begin{table}
\small
\setlength{\tabcolsep}{2pt}
\caption{Population Differences in AI-Related Indices (t-tests)}
\label{tab: T-Test Table}

\centering
\begin{tabular}{lcc@{\hspace{-2cm}}c@{\hspace{-1cm}}c}
\toprule
 & \shortstack{Mean\\General Public} & \shortstack{Mean\\West Point} & \shortstack{Mean\\Difference} & t-statistic \\ 
\midrule
& (N = 7,020) &  (N = 2,360) &  &  \\
\midrule
AI Background Index & 0.25 & 0.41 & -0.16*** & -45.82 \\
AI Experience & 0.29 & 0.38 & -0.09*** & -23.98 \\
AI Familiarity & 0.20 & 0.32 & -0.12*** & -24.92 \\
AI Knowledge & 0.27 & 0.53 & -0.26*** & -41.80 \\
Perceived AI Usage & 0.43 & 0.58 & -0.15*** & -22.09 \\
Programming Experience & 0.34 & 0.42 & -0.07*** & -9.49 \\
STEM Education & 0.36 & 0.59 & -0.23*** & -19.39 \\
AI Beliefs Index & 0.59 & 0.61 & -0.02*** & -5.52 \\

\bottomrule
\multicolumn{4}{l}{\footnotesize Note: Stars denote statistical significance based on two-sample t-tests. *** $p<0.001$, ** $p<0.01$, * $p<0.05$.}\\
\end{tabular}
\end{table}
\end{center}

\begin{figure}
    \centering
    \includegraphics[width=\columnwidth]{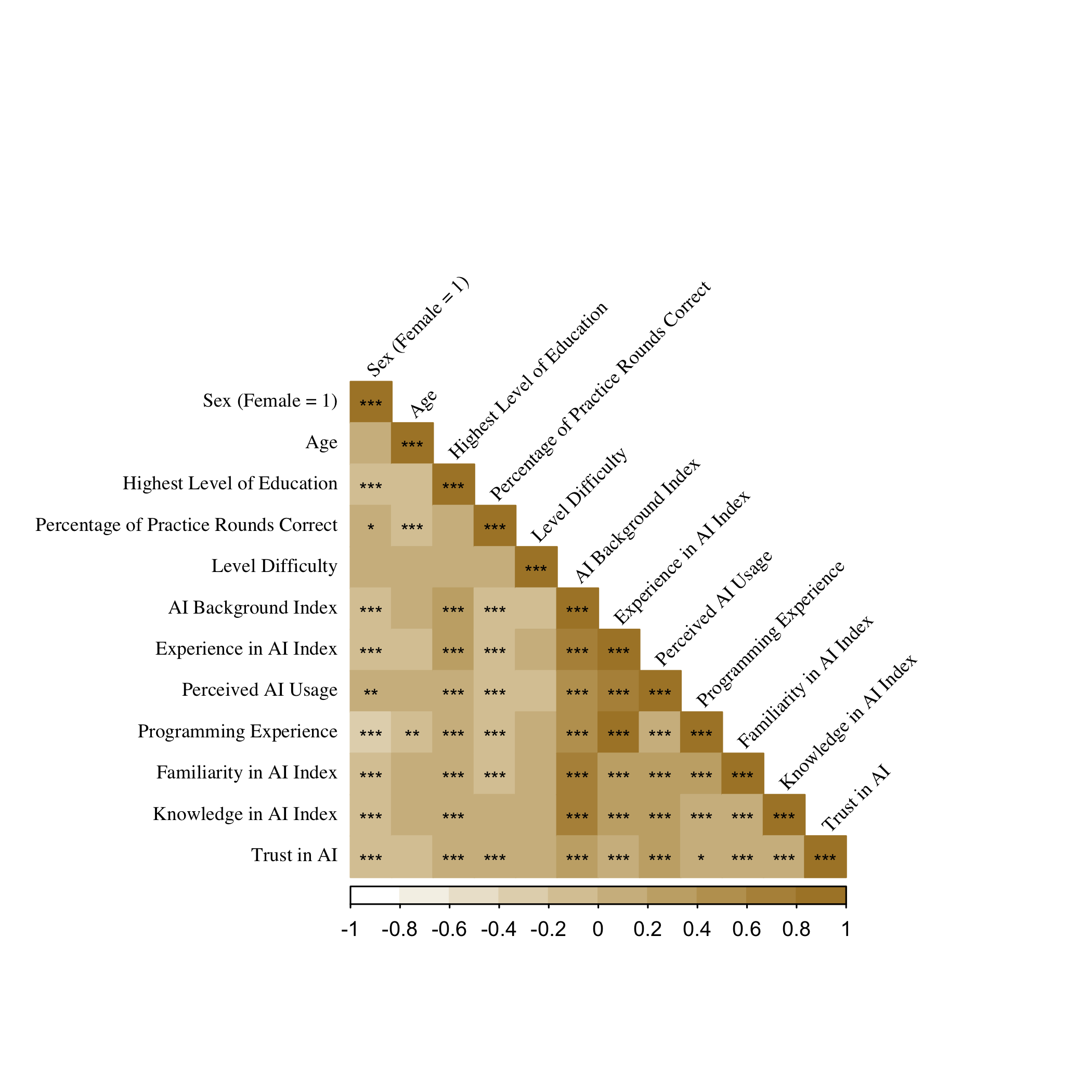}
    \caption{Correlation Matrix General Public}
    \label{fig: Correlation Matrix General Population}
\end{figure}

\begin{figure}
    \centering
    \includegraphics[width=\columnwidth]{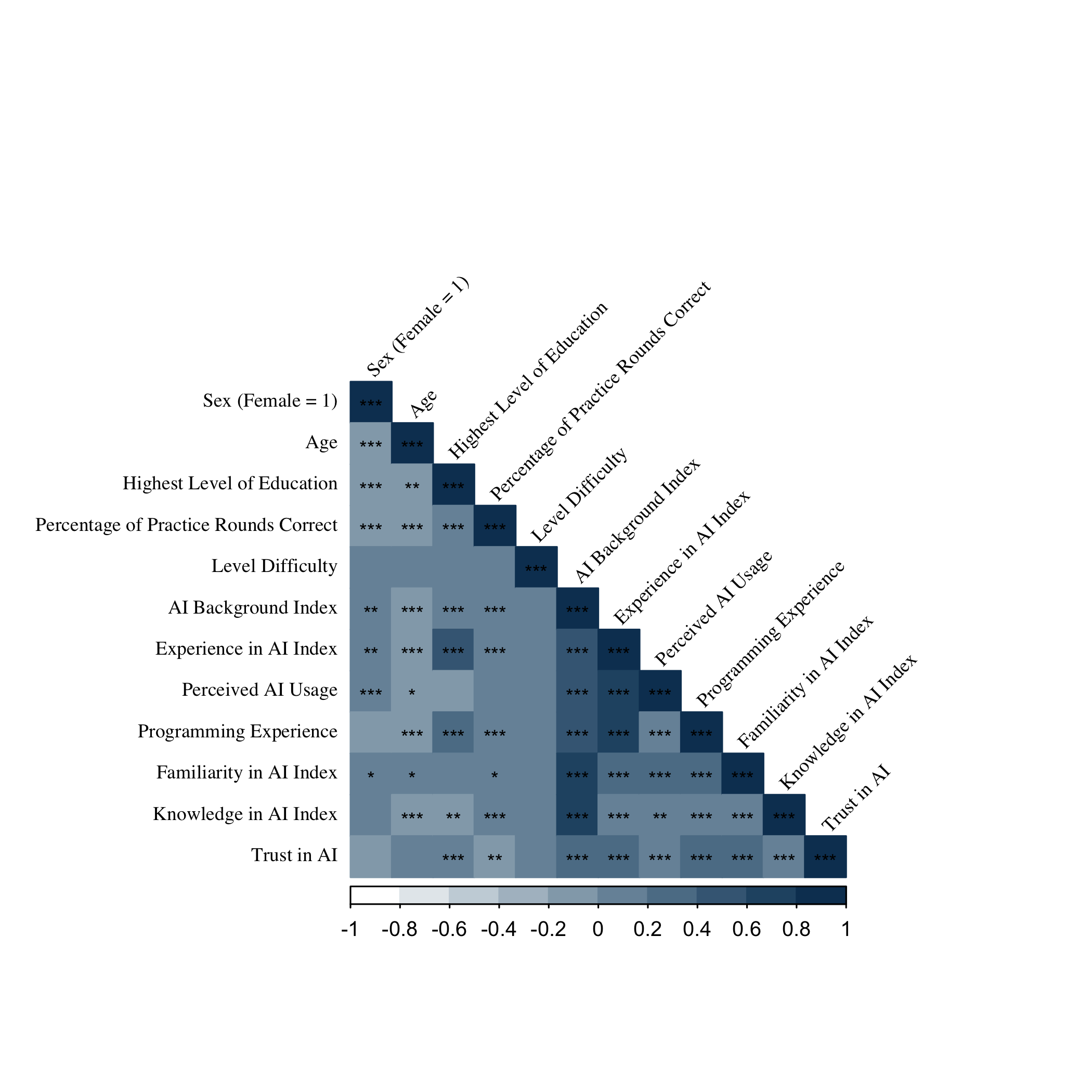}
    \caption{Correlation Matrix West Point}
    \label{fig: Correlation Matrix West Point}
\end{figure}

\begin{figure}
    \centering
    \includegraphics[width=\columnwidth]{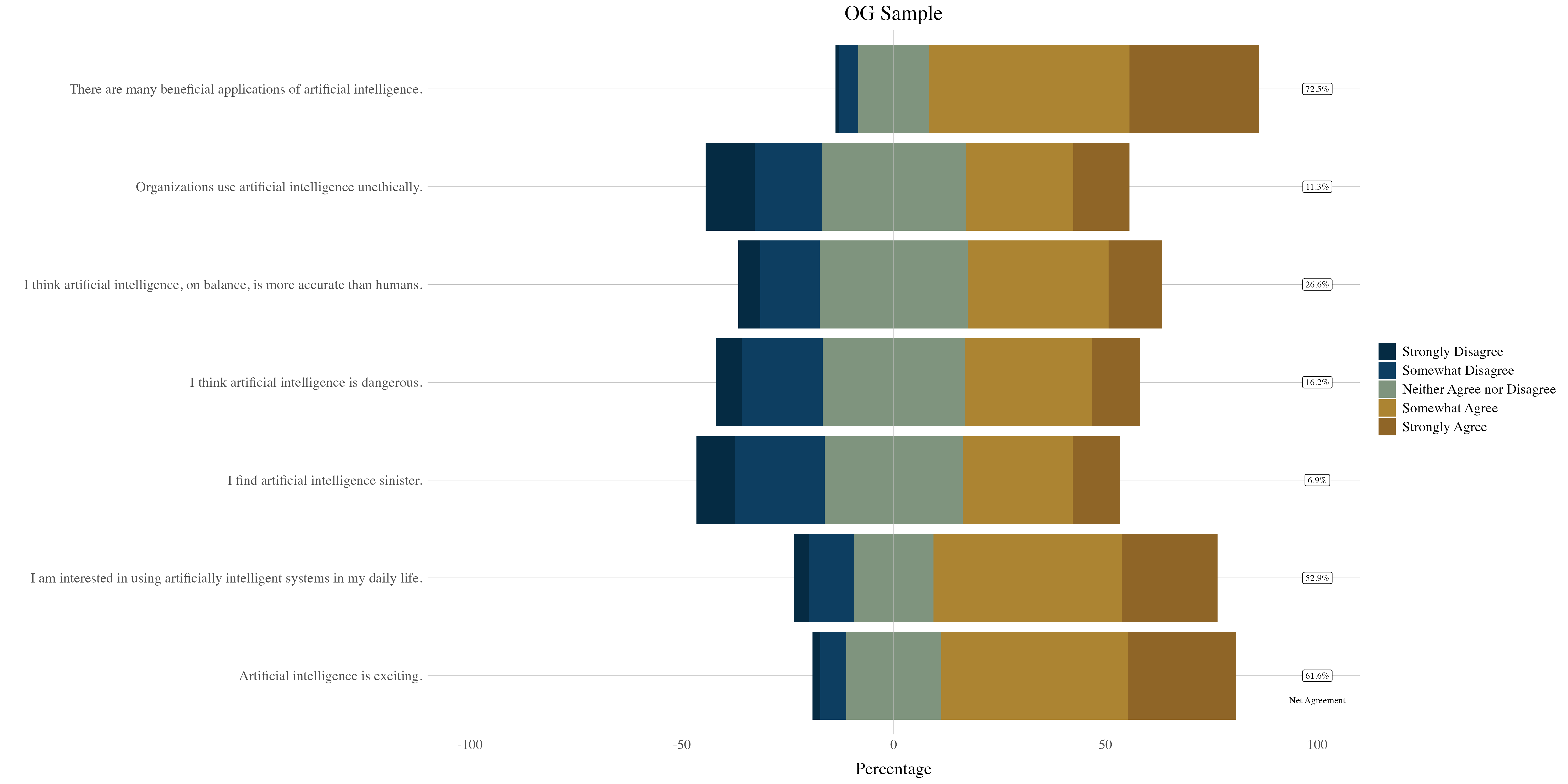}
    \caption{Sentiments General Public}
    \label{fig: Likert General Population}
\end{figure}

\begin{figure}
    \centering
    \includegraphics[width=\columnwidth]{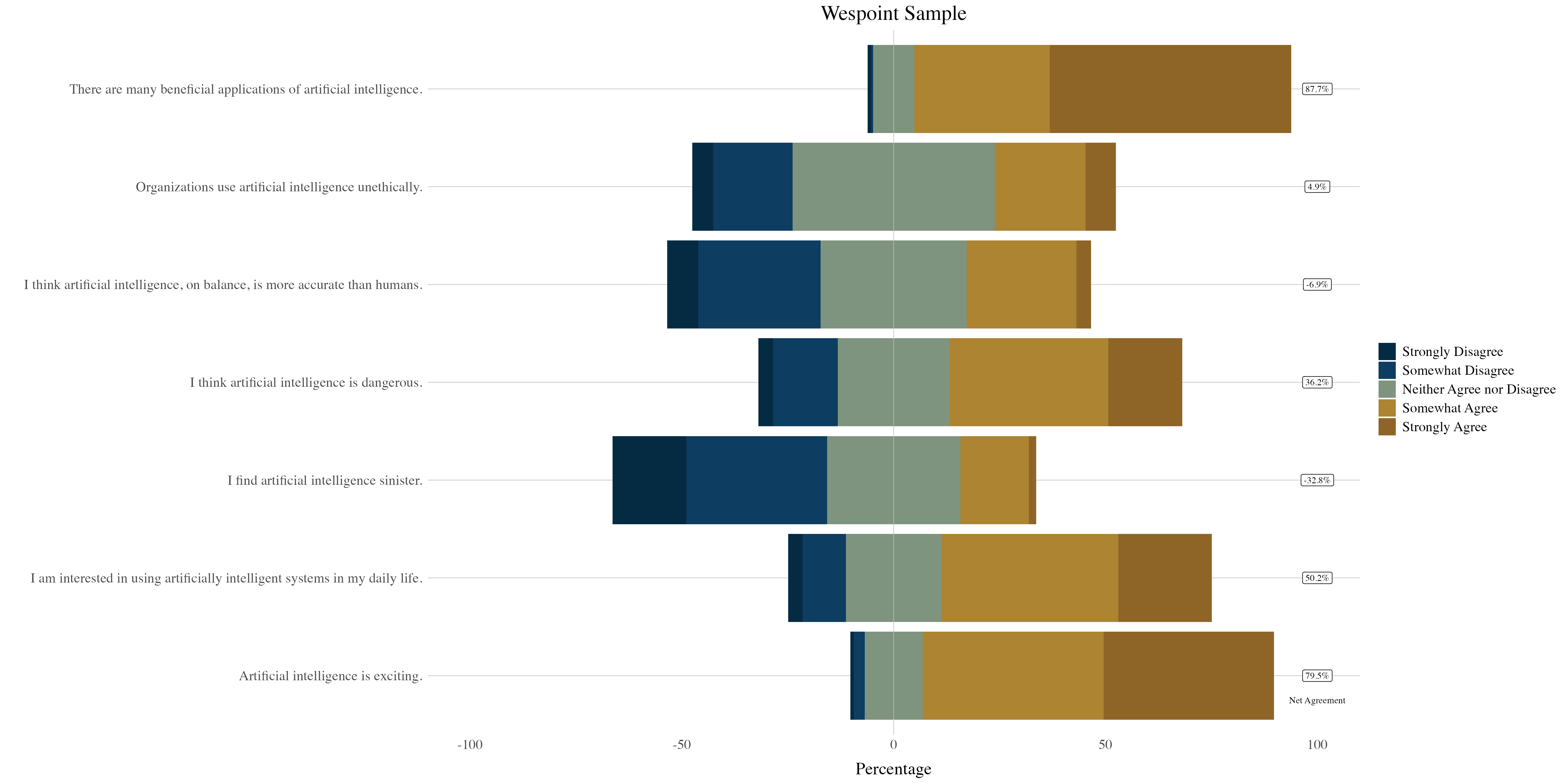}
    \caption{Sentiments West Point}
    \label{fig: Likert West Point}
\end{figure}

\begin{center}
\begin{table}[ht]
\centering
\caption{Covariate-Adjusted OLS Regression Predicting Probability of Automation Bias}
\label{tab: Regression OLS}

\begin{tabular}{l c}
\toprule
 & B/SE \\
\midrule
West Point (vs. General Public) & -0.105** (0.039) \\
AI Beliefs & 0.009 (0.014) \\
AI Background & 0.009 (0.015) \\
Practice Percentage Correct & -0.009 (0.016) \\
Age & 0.017 (0.015) \\
Sex (Female = 1) & 0.009 (0.026) \\
\midrule
Number of Observations & 870 \\
R²	& 0.019\\
Adjusted R²	& 0.012\\
AIC & 695.0 \\
BIC & 733.2 \\
Log Likelihood	& -339.521\\
RMSE & 0.36\\
\bottomrule
\multicolumn{2}{l}{\footnotesize Note: Standard errors in parentheses, clustered at the participant level. ** $p<0.001$, ** $p<0.01$, * $p<0.05$.}\\
\end{tabular}
\end{table}

\end{center}

\end{document}